# A NOVEL CLUSTER VALIDATION APPROACH ON PSO-PAC MECHANISM IN AD HOC NETWORK

S.Thirumurugan[1] and E. George Dharma Prakash Raj[2]

[1]Department of Computer Applications, JJ College of Engineering & Technology,
Trichirapalli, Tamilnadu, India
*s.thiru.gan@gmail.com*
[2]Department of Computer Science and Engineering, Bharathidasan University,
Trichirapalli, Tamilnadu, India
*georgeprakashraj@yahoo.com*

## ABSTRACT

*The ad hoc network places a vital role in contemporary day's communication scenario. This network performance gets up while the clustering phenomenon has been incorporated. The cluster formation using the vital parameters is incredible on deciding the efficiency level of the clustered ad hoc networks. The PSO-PAC mechanism forms clusters based on swarm intelligence by considering energy as crucial parameter. This optimized cluster helps to suits the applications where the energy parameter plays a key role. The clusters formed by this mechanism may not ascertain the compactness of the clusters. Thus, this paper proposes D-PAC as an index based validation mechanism to be handled on clusters formed using PSO-PAC. The cluster formation and validation mechanism have been implemented using OMNET++ simulator.*

## 1. INTRODUCTION

The communication through wireless mode has been an indispensable step in this scenario to suit various applications. This network with ad hoc nature will make the network to be available for various situations. This kind of network comes with the modular growth factor, need to be handled carefully. The clustering gives a solution to this problem. To form the clusters with higher level of efficiency multiple parameters need to be taken into account. It is obviously realized that the distance parameter alone couldn't decide the efficiency of clustering mechanism. Since the nodes which are in wireless network are facing energy drain as a problem with respect to time. Further, the cluster head needs maximum energy among the nodes in a cluster to act as a transceiver. The existing distance based algorithms are showing deficiency in giving an optimum solution. Thus, the multiple parameter based algorithm PSO-PAC(Particle Swarm Optimization of Partitioning Around Clusterhead) has been devised. The swarm intelligence has been the base for the formulation of this efficient clustering mechanism. The behavior of the crow has been considered to devise the procedure of this technique. These optimized clusters need not be compact in their existence. There should be some validation mechanism to analyze the clusters formed. Thus, this study proposes D-PAC(Dunn's Index based Partitioning Around Cluster head) as a cluster validation mechanism to identify the overlapping and separation level of the clusters.

This paper has been organized as follows. Section.1 deals with introduction. Section.2 says about the related works. Section.3 tells about the Ex-PAC mechanism. Section.4 gives out the proposed PSO-PAC procedure. Section.5 deals with the cluster validation mechanism. Section.6 deals with the experimental results. Section.7 brings out the cluster validation results. Section.8 puts down the future works. Section.9 ends up with the concluding remarks.

## 2. RELATED WORKS

The purpose of clustering has been realized when the protocols like AODV[1] has been integrated with the clustering mechanism[2]. Since this addition to the existing protocols produce an improved result. The single parametric algorithm K-means[3] lacks in identifying an efficient cluster. Thus, PAC[4] procedure has been devised to form an efficient clusters in terms of cluster creation time. But PAC has got the down side on leaving many nodes as outliers. This drawback has made to devise another algorithm on top this PAC. The Ex-PAC[5] came out as an extension to PAC which takes entire nodes and produces the maximum possible clusters. The cluster creation process ultimately improved in Ex-PAC procedure. This approach concludes that Ex-PAC has shown significant improvement over k-means in terms of computational speed. The weighted clustering has faced load balancing and global re-clustering as a challenging issue. Thus, IWCA[6] takes relative mobility of nodes into consideration to improve the weighted clustering.

The EWCA[7] takes cluster set up and cluster maintenance phase into the account. The cluster head leaves the role when the energy of the node goes down below the threshold. This will make some other node to become cluster head for the same cluster. The role of cluster head may be switched among the nodes without waiting for the cluster head to get exhausted in their energy. This may balance the energy[8] utilization of the nodes.

The clustering method needs to minimize the energy to set up cluster, maximize the life time of cluster and ensure the stable structure[9]. The power level of nodes decided the sustainability of the cluster head role. The transmission power alone is not sufficient to calculate the weight of the node. The power reward[10] based weight calculation guarantees the uniform power distribution.

The cluster head detection is based on the signal strength[11]. The nodes are also selected based on signal strength between the cluster head and nodes of the respective cluster.

The dominating sets are recognized in the clustered network. In which the minimum independent set[12] can be constructed and the tree structure of the same can be formed later. The connected dominating set algorithm has been a strong procedure to form the clusters.

The SWCA[13] has been proposed to detect the malicious node which acts as cluster head to tamper the degree computation. This also ensures the secured way of transferring the data to the destination with the help of an authentication mechanism.

The NWCA[14] shows an improved weight based algorithms through challenging parameter calculation methodology for weight. The degree computation has been changed to mean connectivity degree. This novel method also considers the energy level of the nodes to play the role of cluster head.

The DWCA[15] considers the cluster formation based on weight, mobility factor and cluster maintenance. This protocol deals with the node addition to the cluster through distinct approach. The weight based clustering has been enhanced[16] on the basis of minimizing the load of the cluster head with the help of threshold value. This limits the cluster size to guarantee the cluster head to last longer.

The aforementioned multi parametric algorithms lack in obtaining an efficient cluster. Thus, The multi-parametric algorithm W-PAC[17] has been devised in order to improve the cluster efficiency in terms of time to form the clusters.

The multi-parametric swarm intelligence based clustering mechanism PSO-PAC[18] takes the necessary parameters to identify the cluster head of the cluster. This parameter optimization will suit specific application.

The IPv4 address no longer will serve the world has been realized. The IPv6 address has got the focus in the present scenario. The configuration of address can be done in two ways. The configuration using DHCPv6 server and stateless autoconfiguration[19]. This configured address can be local link address or global address. This will be decided based on the application requirement. But there must be awareness on the limitations of using the address range has been vital.

The stateless autoconfiguration of IPv6 nodes should have duplicate address detection mechanism. Since when the nodes move across the clusters there is a chance of address duplication. To eliminate this passive duplicate address detection mechanism[20] has been introduced. This method shows better results than passive autoconfiguration for mobile ad hoc networks.

The purpose of autoconfigured address has been recognized while the ad hoc network has come to reality. The limitation of IPv4 and the need of IPv6[21][22] has been understood clearly. Having understood the difference between IPv4 and IPv6, the scenarios will demand the specific way of addressing the nodes. These works are dealing with the autoconfiguration or manual configuration of IPv4 and IPv6 nodes. The clustering as a mechanism comes for this IPv4 or IPv6 autoconfigured node to make the routing simple and also confirm the efficient utilization of bandwidth and resources.

The cluster formation algorithms devised until now missed out on appropriate re-clustering phenomenon. The re-clustering should be based on identifying the strength of the existing clusters. The role of various indices[23] on evaluating the cluster should be understood very well. The cluster classification[24] also plays key role in determining the perfectness of the clusters. Those classifications are of numeric, discrete and partitioned types. It also finds out the preferred clustering method for a given sample set of nodes.

## 3. EX-PAC

The PAC creates the clusters based on Manhattan distance. The Manhattan distance saves time in computing the distance between pair of nodes. The results achieved are not adequate to find out the appropriate cluster in the ad hoc scenario network. This has been further enhanced through Ex-PAC algorithm which has been laid on top of PAC. The experimental results show that Ex-PAC has given better results than K-means algorithm in forming clusters. The formula (1) puts down the calculation of Manhattan distance.

$$\text{Manhattan Distance} = \sum_{i=1}^{n} |x_i - y_i| \qquad (1)$$

### 3.1 Ex-PAC algorithm

1) Assume $N_i$ = temporary Cluster head.
2) Compute Manhattan distance between pair of nodes.
3) if ( MD < range)
   Begin
      Add($N_i$,$C_i$)
      Count = Count + 1
   End
4) Repeat the steps 1 through 3 till all the nodes in the
    cluster are examined.
5) Select the First Cluster which has maximum count value.
6) Select the non cluster nodes.
7) Choose the cluster which includes non cluster nodes.
8) Select cluster returned in step 7 as Second Cluster.
9) Repeat the steps 6,7 and 8 until there is no change in
    Cluster formation.

The Ex-PAC procedure has produced remarkable improvement over the PAC procedure. But still lacks in identifying the perfect cluster head of cluster. The cluster head couldn't be decided based on distance factor only. There should be some mechanism over the Ex-PAC.

## 4. PSO-PAC

The crow behaves noticeably when it finds foodstuff. It alerts neighbours after seeing the eatables through the acoustic signal. Thus, a group of crows will create the cluster. This formation is based on more than one parameter. Those parameters are distance and energy of the crows in communication process. This multi parameter specifies that the cluster formation with the help of swarm intelligence makes the clustering process in a highly efficient way. This behaviour of the crow can be considered in making dynamic clusters.

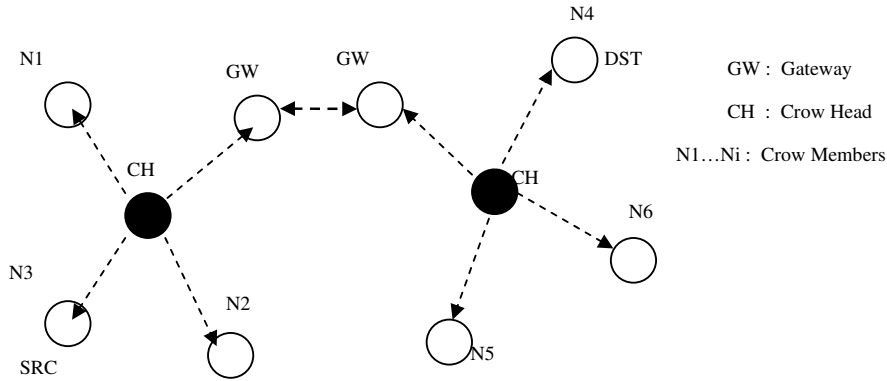

Figure 1. Network Model :Two Clusters

The crow has been taken as a particle in this cluster formation. This particle plays the role of cluster head and members as well. The crow head is the particle which plays cluster head role and crow member is the node belongs to the cluster. The eatables have been considered as a destination to be reached and energy of the crow will be the source of the particle. The activity here is the particle makes the communication after seeing the eatables. Similarly in network communication the source sends the data after knowing the destination through cluster head node. The Fig.1 shows the network model consist of two clusters where each cluster has separate crow head node. The gateway acts as mediator between two clusters. This ensures inter-cluster communication to happen. In this model the destination is far away from the source. This network setup will be retained till the communication between the source and destination gets over. After the data transfer gets over then network has to be reformed with new energy source and destination. In this way the cluster formation happens completely dynamic in nature which suits very well with the behaviour of the crow.

### 4.1 Cluster Formation by Crow Behaviour

1) Input the clusters formed in Ex-PAC procedure.
2) $j = j + 1$
3) $CH = Max\_Energy\_Node(C_j)$
4) Add to $Newcluster(K_j, CH)$
5) $Cluster = C_j$
6) $i = i + 1$
7) $E = Get\_energy(N_i, Cluster)$
8) If ( $E <$ EnergyThreshold )
    $Newcluster ( K_j, N_i )$
9) Repeat the steps 4,5 & 6 until all the nodes belong to the cluster $C_j$ gets examined.
Repeat the steps 2 through 7 until all the clusters of network are examined.
The Max_Energy_Node function returns the node with maximum energy for each cluster. This will be considered as cluster head of the cluster.

### 4.1.1 IPv6 configuration of Clusters

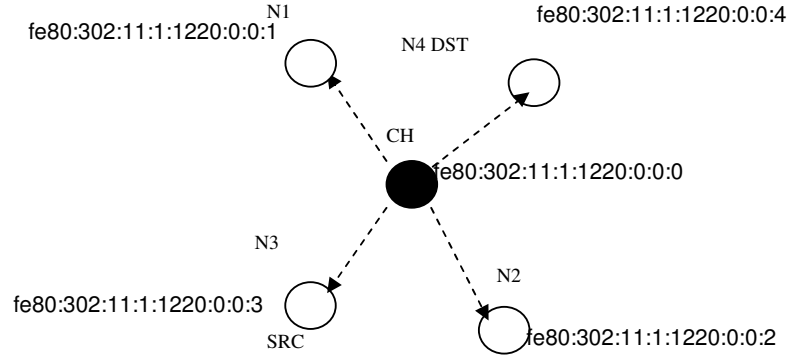

Figure 2. IPv6 Configured Network Model

The cluster formation takes the output of Ex-PAC procedure and recreates the cluster dynamically based on the energy level of nodes. The Fig.2 shows the IPv6 nodes to form the clusters dynamically in a stateful way.

### 4.1.2 IPv6 Configuration Procedure

1) Input the clusters from crow behaviour procedure.
2) Install the DHCPv6 server process in Cluster head node.
3) Cluster head sends hello message to all the members.
4) For each Cluster Ci
   Begin
     For each member node Nj
       Begin
         Node Nj sends reply to the Cluster head.
         Cluster head sends an IPv6 address to Nj.
       End
   End
5) Repeat the step 4 for j = 0 ... Number of Nodes in cluster Ci and for i = 0 to Number of clusters.

The IPv6 configuration procedure configures the nodes belong to the clusters under stateful way using DHCPv6 server. In this method the cluster head sends hello message to all the members at the initial stage. The member nodes send the reply to the cluster head. This reply informs the member belong to the cluster where the cluster head exists. The cluster head now sends IPv6 address to the member node as a part of configuration.

## 5. D-PAC : CLUSTER VALIDATION

This parameter helps to validate the clusters formed using PSO-PAC procedure. This validation process will identify the clusters compactness and the separation of the clusters based on the parameter value computed. The formula (2) shows that the min distance between the clusters Cp and Cq. This could be obtained by taking all the nodes belong to the two clusters to apply the distance computation such as Manhattan distance.

$$dis(Cp, Cq) = \min_{m \in Cp, n \in Cq}(d(m, n)) \qquad (2)$$

In the formula (3) the max diameter of the cluster will be identified with the help of the pair of nodes which are placed far apart within the cluster. This process takes all the nodes within the cluster.

$$\text{dia}(C_p) = \underset{x,y \in C_p}{\text{Max}}(d(x,y)) \qquad (3)$$

The formula (4) gives out the dunn's index value which is ratio between the minimum distance of clusters and maximum diameter of cluster.

$$\text{Min}_{p=1..nc} \left\{ \text{Min}_{q=i+1..nc} \left[ \frac{\text{dis}(C_p, C_q)}{\text{Max}_{r=1...nc} \text{dia}(C_r)} \right] \right\} \qquad (4)$$

The dunn's index value calculated using the above mentioned formula tells how the clusters are segregated with each other and also represents the strength of the cluster based on the nodes exist.

### 5.1 D-PAC Algorithm

1. Create the clusters using the PSO-PAC procedure.
2. Find the distance between pair of clusters Cp and Cq.

$$d(m,n) = \sum_{i=1}^{N} |m_i - n_i|$$

   Dist(Cp,Cq) = min(d(m,n)) , m∈Cp , n∈Cq
3. Repeat the step 2 for all the pairs of clusters.
4. Find the maximum diameter of the cluster.
       Diameter(C) = max(d(m,n)) , m&n∈C
5. Repeat the step 4 for 1 to N clusters.
6. Calculate Dunn's Index.
     Dunn's index = Min(dist(Cp,Cq)) / Max(diameter(Cr))
7. if Dunn's index between 0 and 0.5 then
         Clusters are compact and separation is less.
      Else
        if Dunn's index between 0.51 and 1.0
            Clusters are compact and well separated.

This algorithm takes PSO-PAC output as input clusters for the validation process. The Dunn's index has been calculated based on the formulas aforementioned. The resultant value will reveal that the clusters are overlapping or well separated.

### 6. EXPERIMENTAL RESULTS

This PSO-PAC algorithm and D-PAC validation mechanism has been implemented in OMNET++ and the results are tabulated. This work has been carried out with the system configuration of 64bit AMD processor, 2GB RAM and windows XP as an operation system. This simulation has been done for 25 nodes and 50 nodes.

Table 1. Simulation Parameters

| Parameter | Values |
|---|---|
| N (Number of Nodes) | 25, 50, 300 |
| Space (area) | 100 ×100 |
| Tr (Transmission range) | 20m |
| Execution Time | 5 sec |
| Threshold(Energy) in Units | 500 |

The Table.1 shows the simulation parameters considered while simulating the study.

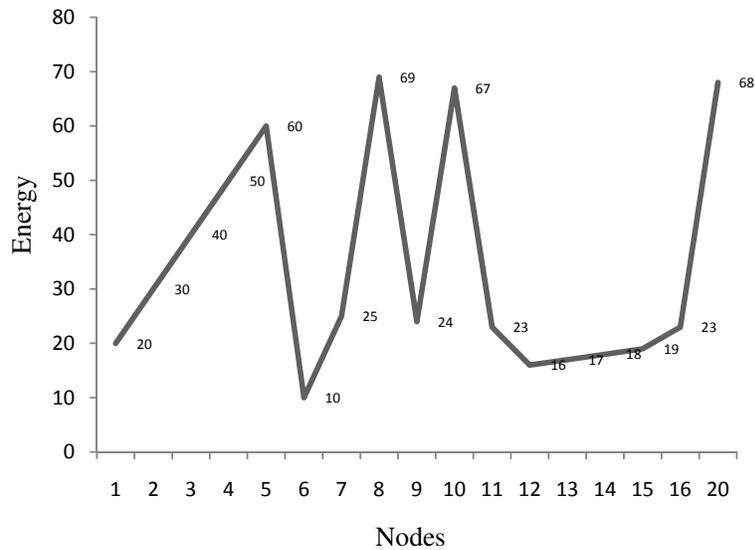

Figure 3. Energy Graph for 25 Nodes : Cluster C1

The cluster heads could be fixed and need to be changed in accordance with the time T. The energy level graph Fig.3 at a specific time T identifies the node which is eligible to be considered as cluster head at the time period T1. After time period T1 some other node will be chosen as cluster head at time period T2. This graph clearly reveals the cluster heads selection in the order.

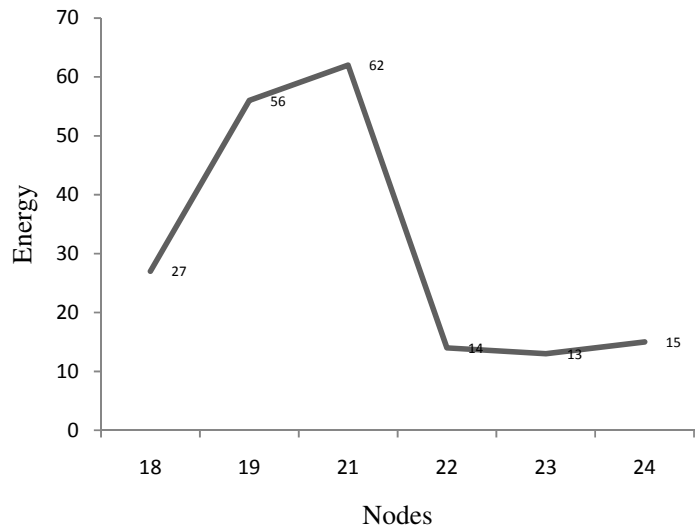

Figure 4. Energy Graph for 25 Nodes : Cluster C2

The Fig.4 shows the nodes and their respective energy level belongs to cluster C2. This ushers that the node21 will be chosen as cluster head since it has high energy level at time T1.

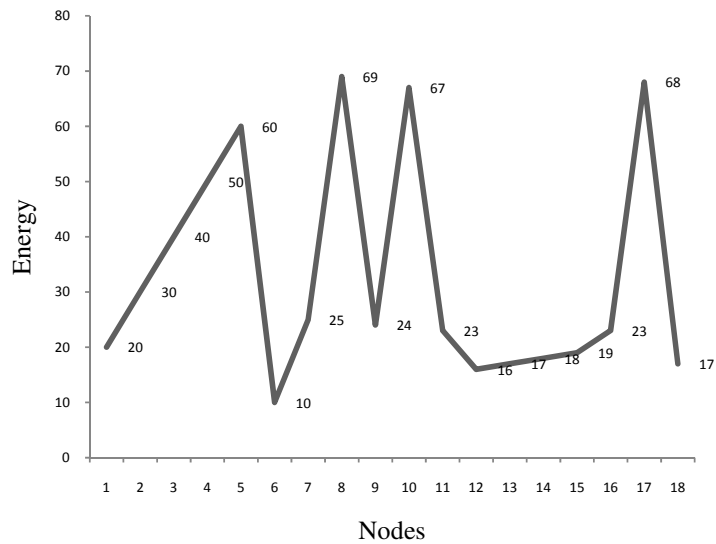

Figure 5. Energy Results for 50 Nodes : Cluster C1

The Fig.5 ushers the energy based results for the sample size of 50 nodes. This shows that node8 takes the role of cluster head.

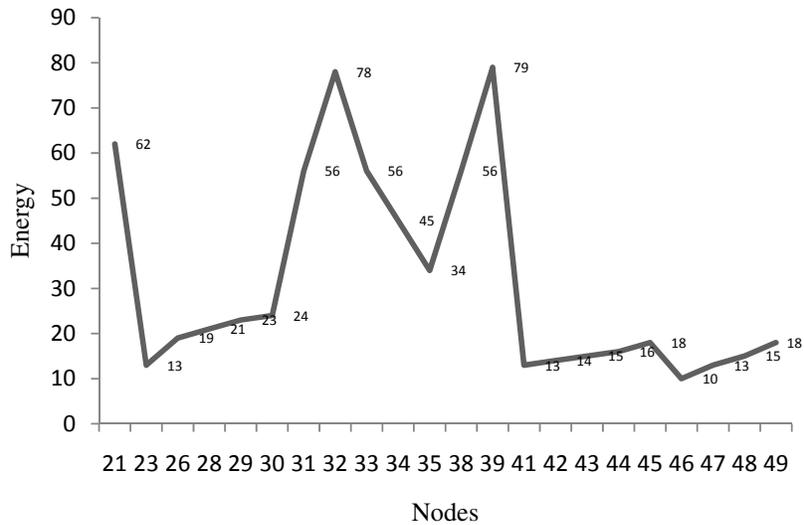

Figure 6. Energy Results for 50 Nodes : Cluster C2

The Fig.6 ushers the energy based results for the cluster2 where the sample size has been 50 nodes. This identifies node39 as the cluster head.

## 7. D-PAC : CLUSTER VALIDATION RESULTS

The clusters formed using PSO-PAC will not ascertain the maintenance of the clusters. There should be some mechanism to analyze the clusters formed to confirm the compactness of the clusters. This tells the clusters separation and overlapping percentage.

Table 2. Results : D-PAC

| Num. Of Nodes | Dunn's Index | Separation | Overlap | Compact |
| --- | --- | --- | --- | --- |
| 25 | 0.52 | 52% | 48% | High |
| 50 | 0.48 | 48% | 52% | Low |
| 300 | 0.01 | 10% | 90% | Very Low |

The Table.2 shows the results obtained in a tabulated manner. It is clearly reveals that the value of index indicates the segregation of clusters and also the level of cluster overlap. The calculation gives 0.52 as the Dunn's index result for 25 nodes. This value indicates the separation of clusters. The clusters are 52% apart has been identified through this calculation. This drives to understand that there is a chance of 48% overlap in clusters of network. The index value for 50 nodes 0.48 has been obtained says that the clusters separation is less than that of 25 nodes. This proportionately increases the overlapping of the clusters which is slightly higher when compared with 25 nodes. The compactness is decided based on the intra cluster relationship of nodes. The separation is high which indicates the high intra cluster relationship ends up at high compactness. The compactness for 50 nodes has been low when compared with the 25 nodes. The compactness for 300 nodes has been very less to indicate the closeness of clusters. This will increase the interference between the clusters. This can be solved by invoking re-clustering process.

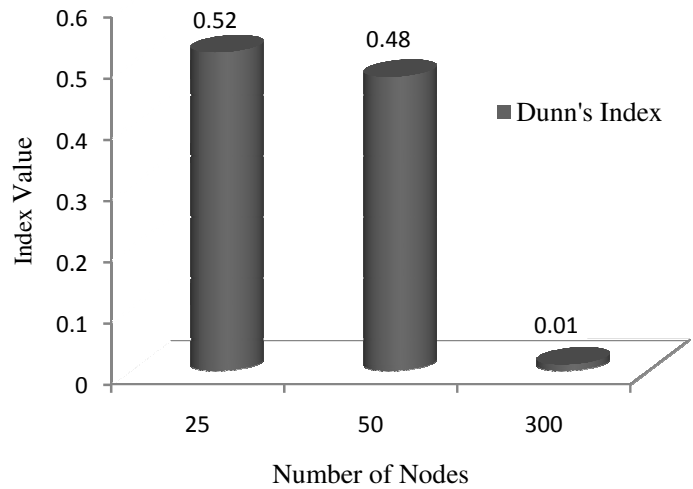

Fig.7 Graphical illustration

The Fig.7 shows the graph of Nodes against Dunn's index. This illustration makes to understand the index value for the different sample size. The higher the value of index the stronger the cluster and separation would be high.

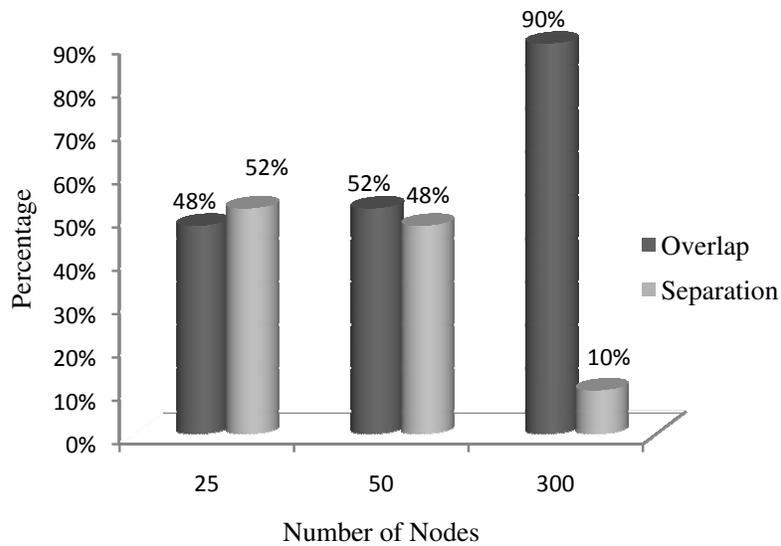

Fig.8 Graphical illustration

The Fig.8 tells the possibility of overlapping based on the separation which in term computed based on the Dunn's index. This graph makes to visualize the occurrence of overlapping clusters in the selected sample set of nodes.

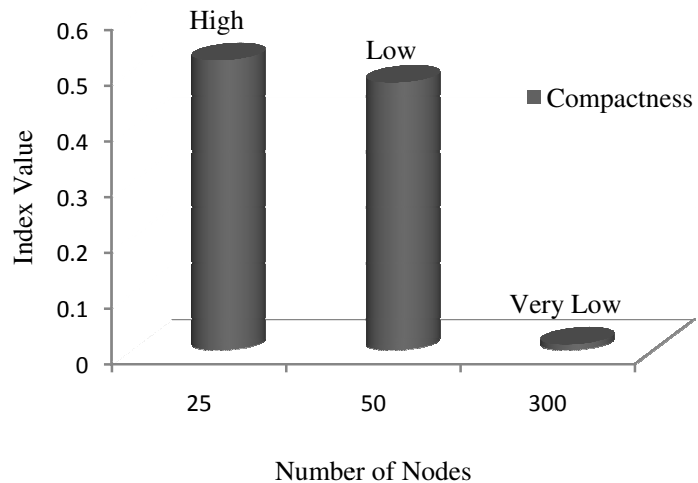

Fig.9 Graphical illustration

The Fig.9 identifies the compactness level of the clusters. This shows that the smaller nodes sample has high compactness than larger one.

## 8. FUTURE DIRECTIONS

This work has considered dunn's index as a validation parameter on clusters. This work should be done for various validation parameters to identify the strength of clusters.

## 9. CONCLUSION

This study has devised the clusters using PSO-PAC algorithm. These clusters are validated for their compactness using D-PAC procedure. This D-PAC result obviously determines the closeness of the clusters. This closeness helps to decide about the clusters which are interfering due to overlapping. In this approach intra-cluster strength has been measured by separation percentage. The higher separation level makes the understanding that the bandwidth could be reused with less or without interference. The periodical validation of these measures tells the time of re-clustering the network. The implementation using OMNET++ confirms the clusters separation level with the help of the Dunn's index value.

**Authors**

**Mr.S.Thirumurugan** completed his Masters Degree in Computer Applications and Master of Philosophy in Computer Science. He has around 9yrs of experience in teaching field which includes his association with the research work. He has published his work in two international journals, presented four papers at the international level and also two papers at the national level. His area of research work falls on Ad hoc networks and their applications on real world scenario.

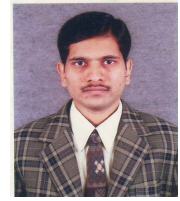

**Dr.E.George Dharma Prakash Raj** completed his Masters Degree in Computer Science and Master of Philosophy in Computer Science in the years 1990 and 1998. He has also completed his Doctorate in Computer Science in the year 2008. He has around twenty-one years of Academic experience and thirteen years of Research experience in the field of Computer Science. Currently he is working as an Assistant Professor in the Department of Computer Science and Engineering at Bharathidasan University, Trichy, India. He has published several papers in International Journals and Conferences related to Computer Science and has been an Editorial Board Member, Reviewer and International Programme Committee Member in many International Journals and Conferences.

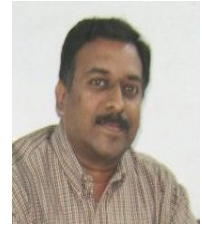